\documentstyle[12pt,aasms4]{article}

\begin{document}
\title{$K$-Band Spectroscopy of (Pre-)Cataclysmic Variables: Are Some Donor Stars Really
Carbon Poor?}

\author{Steve B. Howell\altaffilmark{1,4}}
\author{Thomas E. Harrison\altaffilmark{2,4}}
\author{Paula Szkody\altaffilmark{3,4}}
\and
\author{Nicole M. Silvestri\altaffilmark{3}}

\altaffiltext{1}{NOAO, 950 N. Cherry Ave., Tucson, AZ 85719:
(E-mail: howell@noao.edu)}
\altaffiltext{2}{New Mexico State University, P.O. Box 30001, Las Cruces, 
NM 88003: 
(E-mail: tharriso@nmsu.edu)}
\altaffiltext{3}{University of Washington:
(E-mail: nms/szkody@washington.astro.edu)}
\altaffiltext{4}{Visiting observer Keck II telescope}

\begin{abstract}
We present a new sample of $K$-band spectral observations for CVs: non-magnetic and
magnetic as well as present day and pre CVs. The purpose of this diverse sample is to 
address the recent claim that the secondary stars in dwarf novae are carbon deficient,
having become so through a far more evolved evolution than the current paradigm predicts.
Our new observations, along with previous literature results, span a wide range of orbital 
period and CV type. In general, dwarf novae in which the secondary star is seen show weak to no CO
absorption while polar and pre-CV donor stars appear to have normal CO
absorption for their spectral type. However, this is not universal.
The presence of normal looking CO absorption in 
the dwarf nova SS Aur and the hibernating CV QS Vir and a complete lack
of CO absorption in the long period polar V1309 Ori cloud the issue.
A summary of the literature pointing to non-solar abundances including enhanced NV/CIV
ratios is presented.
It appears that some CVs have non-solar
abundance material accreting onto the white dwarf suggesting an evolved secondary star
while for others CO emission in the accretion disk may play a role. 
However, the exact mechanism or combination of factors causing the CO absorption anomaly 
in CVs is not yet clear.
\end{abstract}


\section{Introduction}

Cataclysmic variables are semi-detached binaries in which a white dwarf primary
is accreting material from its close neighbor, a low-mass Roche-lobe filling
object. The donor stars are often taken to be of the main sequence variety, but
observational evidence often suggests that they are more evolved.
Close binary star evolution has some known basic properties (e.g., gravitational radiation)
but the details of the ``standard model" are not yet fully in hand. Additionally, some 
``oddball" systems (e.g., Eggleton 2006) are known and generally swept under the rug.

Over the past few years, infrared spectroscopy has revealed a new factor
to be dealt with in our understanding of binary evolution - the apparent carbon
deficiency or low
carbon abundance observed in the secondary stars of some dwarf novae type 
cataclysmic variables (Harrison et al., 2004). 
To make things even more interesting, it appears that the weakness of carbon (as
evidenced by very weak or missing CO absorption in the spectra) is only mainly 
an issue for cataclysmic variables (CVs) 
with non-magnetic white dwarf (WD) primary stars. The highly magnetic WD
primary systems, the polars, have secondary stars that appear to be completely
normal (Harrison et al., 2005).

The generally accepted present-day paradigm for CV evolution (Howell et al.,
2001) cannot account for the low carbon abundance observed 
or for a different evolutionary sequence for magnetic CVs vs. non-magnetic CVs.
Evolution models such as those by Beuermann et al. (1998) and Barraffe and Kolb (2000)
explore donor stars which have significantly evolved prior to the start of 
mass transfer. Marks and Sarna (1998) explore the chemical evolution expected in
secondary star systems which undergo nova explosions and which go through a 
common envelope stage.
 
Gansicke et al. (2003) note that eight CVs are known to have large NV/CIV emission line
ratios as seen in their UV spectra. Cheng et al. (1997) found that the TOAD WZ Sge was over
abundant in carbon (5x) as well as in nitrogen (3x) with respect to solar.
Thus, it is not just the weak or absent CO observed in the $K$-band that seems to be implying
that some systems have non-solar abundances, generally indicative of more evolved stars
in which CNO burning has occurred.
To date, there has been no detailed elemental 
abundance study of CV donor stars, a useful enterprise to undertake
in order to provide metalicity values and 
evolution indicators such as the 12C to 13C ratio. 

We present new IR spectroscopy of a number of CVs and related non-interacting 
pre-CV binaries.
Using these new results, along with those previously published in the literature,
 allows us to span a
wide range in orbital period and CV subtype. We can then  formulate a picture
of the nature of the secondary stars as a function of orbital period,
interaction state, and magnetic field of the primary white dwarf. We
attempt to use the observations to constrain the carbon abundance as
it relates to binary evolution.

\section{Infrared Spectroscopic Observations}

All our infrared spectra presented here were obtained at the Keck II telescope
on Mauna Kea, Hawaii on 4 and 5 March 2007 UT 17 using the NIRSPEC.
NIRSPEC was used in low resolution mode with the slit width set to
0.38" as we had seeing near 0.4" on both nights. 
We chose a grating tilt to cover the $K$-band spectral range of 
2.04-2.40 micron with a dispersion of 4.27 \AA/pixel or a velocity resolution
near 110 km/sec. All stars were observed using
a standard ABBA nod pattern on the slit and the four nod exposures were
combined into final spectra depending on the S/N per nod. 
Our observations were not all obtained under photometric conditions and the use of a narrow
slit both preclude assignment of true fluxes to the spectra. 

Observations of bright A-type stars nearby in
time and location were used for telluric correction. 
Spectrophotometric stars were observed with the same setup and we estimate our fluxes are
accurate to $\sim$15\%. 
The Keck spectra were reduced using the IDL routine REDSPEC especially developed
for NIRSPEC reductions. A stars are nearly featureless in the $K$-band except for
Brackett$\gamma$ which was removed by interpolation across the continuum
on both sides of the line during reductions.
A complete description of our reduction process is presented in Harrison et al.
(2004) and Table 1 presents a log of the IR observations.

\section{Analysis}

We present results for each star separately below, 
organized by their orbital period; long to short.
Late-type dwarf templates, near the observed donor star 
(or expected donor star) spectral type, are shown in some figures for comparison purposes and all are
from the IRTF spectral library (Rayner et al., 2009)\footnote{The IRTF spectral
library is available in digital form at 
http://irtfweb.ifa.hawaii.edu/~spex/IRTF\_Spectral\_Library/index.html.}.
For all the template stars from the IRTF spectral library, fluxes are given in 
W m$^{-2}$ $\mu$m$^{-1}$ while our Keck spectra are presented in normalized flux units of the same kind.

For most of the systems listed
below, we estimate the spectral type of the secondary star. These spectral
types are based on the similarities in strengths of the strongest absorption
features to MK template spectra. As with our previous efforts, however, 
unusual line strengths can be present which confuse the precise classification,
and any such spectral types are probably only good to $\pm$ one subtype for the 
best of the spectra. The emperical method we use 
is discussed in detail in Harrison et al. (2004, 2005b).
Table 2 summarizes our findings for the present data set.
We will abbreviate SDSS sources in the paper as follows:
SDSS0830 (=SDSSJ083038.80+470246.97), SDSS0757 (=SDSSJ075721.06+323054.39), 
SDSS0743 (=SDSSJ074301.92+410655.18), and SDSS0837 (=SDSSJ083751.00+383012.5).

\subsection{HS1136+6646}

HS1136+6646 was identified as a young, post common envelope DAO white
dwarf binary, that is a pre-CV, by  Sing et al. (2004). 
Velocity analysis of optical spectra of HS1136 revealed a hot WD and narrow
emission lines assumed to be from an irradiated secondary. Using this
information, Sing et al. inferred a likely M2.5V secondary star 
in a 20 hour orbit.
Later, Liebert et al. (2006) discovered a third body in the system at
a separation of 1.3 arcsec with a position angle of 54.5 deg, which was likely
a common proper motion object. 

When we pointed the Keck II telescope at HS1136, the acquisition imager (SCAM)
easily revealed two sources differing in brightness by 1.9 mag with the pre-CV 
(SW component) being fainter than the K7V (NE) component. 
We measured a separation of 1.2" at a position angle of $\sim$52 degrees,
consistent with the values found by Liebert et al. (2006).

We obtained a $K$-band spectrum of both sources, the CV and the K star
(Figure 1).  Narrow emission lines due to He (2.056 microns) and
Brackett$\gamma$ are seen in the pre-CV spectrum. Sing et al. (2004) attributed the optical
emission lines to irradiation of the secondary (by the 70,000K WD) and noted that only 
the He II $\lambda$4686 emission line moved with the white dwarf velocity.
However, they found from high resolution spectroscopy that H$\alpha$ shows
a double-peaked structure (velocity separation = $\sim$90 km/sec)
with a stronger narrow component, suggestive
of the presence of accretion disk emission in addition to the irradiation lines.
The HeII $\lambda$4686 emission 
line gave K1=69+/-2 km/s, and an origin from the white
dwarf was confirmed as this line had similar velocity amplitude and phase
to WD lines measured from phase-resolved FUSE observations (Sing et al. 2004).

We see no absorption features from the M2.5V secondary in our $K$-band spectrum of the pre-CV.
This is not surprising as the hot (70,000 K) white dwarf continuum dominates
the system flux even into the $K$-band, and the secondary was only observed in the optical
due to H emission lines produced by strong irradiation.

The NE component, the common proper motion K7V star, 
has an IR spectrum that presents Mg I (2.106 microns),
Al I (2.116 microns), Na I (2.21 \& 2.35 microns), 
Ca I (2.26 microns), and CO in absorption. The $K$-band spectrum of this object is contaminated by the
closeness of the CV, and in particular we see in Fig. 1 some residual Br$\gamma$ emission from the
pre-CV. 

\subsection{V471 Tau}

This famous Hyades member 
binary (Ibanoglu et al., 1994; Rottler et al., 2002; and references
therein) was observed as a ``standard" pre-CV star. The 0.52 day binary consists
of a white dwarf and a K2V secondary star, the latter being highly active.
O'Brien et al. (2001) discuss oddities in both stars, the white dwarf being too
young and too hot for its mass as a cluster member and the K2V dwarf being too
large compared with other similar mass K2 dwarfs in the Hyades cluster.
V471 Tau is often cited as the canonical pre-CV.
Dhillon et al. (2002) refute the earlier claim by Dhillon \& Marsh (1995) and
Catalan et al. (2001) that there is anything unusual about the 13CO feature
at 2.345 microns in V471 Tau. 

We find that our IR spectrum of V471 Tau agrees well
with a K2$\pm$1 main sequence star being similar to the two templates
shown in Figure 2. This classification agrees with the K2V choice 
of Dhillon et al. (2002) using
the 13CO feature. We find no absorption features that are
inconsistent with normal solar abundances and the CO bands are present and 
with the strengths expected for an early KV star.

Assuming a 100\% contribution from a normal K2 V star (M$_K$=
7.3) and convolving a K filter with our spectrum (giving K=10.6), we find a
distance to V471 Tau of 47.4 pc, in agreement with the HIPPARCOS parallax value
of 47.6 pc (de Bruijne, Hoogerwerf, \& de Zeeuw, 2001).
Other than the slight oddities noted in the literature, the secondary star 
in V471 Tau appears to be fairly normal.

\subsection{BT Mon}

BT Mon is an old nova (Nova Mon 1939) with an orbital period near 1/3 of a day and
with deep (2.7 mag) eclipses. Currently it is classified as a dwarf nova or a novalike
(Warner 1995).
At 15th magnitude in V, this binary has not been studied extensively
but two comprehensive optical spectroscopic works have been presented
(White et al., 1996; Smith et al., 1998). White et al. found inconclusive evidence
to support an accretion disk in BT Mon but suggested a possible magnetic white dwarf
exists in the binary. Smith et al. detected weak
absorption star features from the secondary during eclipse. Comparison with
template spectra provided them with a best fit at G8V, but this was 
the same spectral class as the earliest template they used so an earlier type donor star
may be possible. Using the relation determined by Smith and Dhillon (1998), BT Mon's secondary
star should be near K3V. 

$K$-band spectroscopy, presented in Figure 3, shows weak HeI emission and a strong,
narrow Brackett$\gamma$ line. 
The blue continuum and the strong emission lines 
argue for a strong disk contribution in BT
Mon, a typical result of the high mass transfer rate of old novae. 
Thus, any secondary star contributions are outshone by the accretion disk flux.
It is not surprising that such a weak spectral contribution
lends itself to detection in the optical only during
the deep eclipse when the disk light is vastly diminished.

\subsection{V1309 Ori}

V1309 Ori (RX J0515.6+0105) is listed as an eclipsing magnetic CV 
with the longest known orbital period for a polar (7.98 hr) and 
an M0 secondary star
(Garnavich et al. (1994). 
These authors note that a normal M0V is not large enough to fill the Roche
lobe in such a long period system. A low resolution infrared spectrum was
presented by Harrop-Allin et al. (1997) that revealed H I and He I emission, but
no features that they could attribute to the secondary star. They suggested that
the IR spectrum was 
dominated by cyclotron emission from the B=33-55 MG white dwarf primary.

Our Keck spectrum is presented in Figure 4 and covers 0.04 in orbital phase. 
The Brackett$\gamma$ emission line is seen as well as He 2.056 
micron emission. 
%The H emission line appears to be too narrow to come from an
%accretion disk in this high inclination eclipsing binary.
The H emission line is suggested to be from the
gas stream and accretion column of material falling toward the white dwarf
magnetic pole. We see no evidence for cyclotron humps, but given 
the apparent strong B field none would be expected in this wavelength region. The 
rising blue continuum is consistent with an origin in the Rayleigh-Jeans tail of
cyclotron emission; a similar conclusion was reached by Harrop-Allin et al. (1997).
However, the rising continuum is also
consistent with the spectral energy distribution in this region for K and early M stars.

We clearly detect secondary star
features as absorption lines due to Na I, Ca I, and Mg I. Figure 4 compares our IR
spectrum to single star templates covering K0V to M0V with a best fit (based on line
ratios) of K6-K8. No CO absorption is
detected from the secondary star in V1309 Ori although it should be for this temperature
(spectral type) as can be seen in the template spectra.
For an orbital period of 8 hr, the secondary star would be expected to
be a G3V ((Smith \& Dhillon 1998) and modeled to have a mass of 1.19 M$_{\odot}$ 
and a radius of 0.98 R$_{\odot}$ (Howell et al., 2001).
Our $K$-band spectrum seems to clearly eliminate the possibility that the secondary is
such an early G star and we suggest that the donor is indeed near K7 in spectral type but
is a subgiant star as has been shown to exist in the similar orbital 
period magnetic CV AE Aqr (Harrison et al., 2004a).

V1309 Ori presents a mystery given its lack of CO absorption. 
All other polars have shown
normal abundance secondary stars with CO absorption in their $K$-band spectra
(Harrison et al. 2005) while
a lack of CO has been observed in most dwarf novae. 
BY Cam does show an enhanced NV/CIV ratio, however, suggestive of
CNO processed material accreted from the secondary (Bonnet-Bidaud \& Mouchet,
1987) but no $K$-band spectrum of its donor star has yet been obtained.
AE Aqr, the odd intermediate polar, shows an enhanced NV/CIV ratio as well as 
very weak CO (Harrison et al. 2007).
Perhaps V1309 Ori shares some property with AE Aqr or BY Cam
leading to its lack of CO absorption.
It would be hard to reconcile that V1309 Ori contains a magnetic field strength similar to 
IPs given 
the cyclotron spectrum observed and modeled by Garnavich et al. (1994).

\subsection{SS Aur}

SS Aur is a fairly normal dwarf nova with an orbital period of 4.38 hours, 
therefore above the CV period gap. The binary inclination (38 degrees) and hot white dwarf
primary (30,000K: Lake \& Sion 2001) both tend to produce a bright, rising blue continuum.
Our $K$-band spectrum (Figure 5) shows the blue continuum rise as well as broad, 
double-peaked emission lines due to Br$\gamma$ and He I. 
Absorption features from the secondary star are also easily observed in Figure
5 and agree with a M4V star, not with the expected secondary type of
late K to early M (Smith and Dhillon 1998). We also
note that Na I and Ca I are a bit too strong for a M4V while 
the spectrum contains completely normal looking CO absorption bands (for the observed 
secondary type). 
SS Aur is one of the rare DN which have relatively normal CO.

The $K$-band spectrum shown in Figure 5 is consistent with solar abundances in that
the line ratios agree with our M4V template while the bluer slope herolds the
the presence of a hot white dwarf and accretion disk in the SS Aur system.
According to AAVSO records our spectrum was obtained in quiescence, 10-12 days prior to a normal
outburst. SS Aur contains one of the hottest white dwarfs known in a dwarf nova due 
to its high mass transfer
rate estimated at 10$^{-10}$M$_{\odot}$ yr$^{-1}$ (Lake and Sion 2001). This mass transfer
rate is very close to the critical rate for a solar mass white
dwarf such that DN outbursts would not occur. This high mass transfer rate may indicate
that SS Aur is a relatively new CV, just formed above the period gap 
(see Howell et al. 2001).

It is interesting to compare SS Aur with U Gem. These two DN have similar orbital periods
yet their CO behavior is quite different (c.f., Harrison et al., 2004). 
SS Aur is a high mass transfer system, thus
we would expect an extensive accretion disk. 
Completely normal C abundance has been noted for this object (Godon et al. 2008)
and we see normal looking CO absorption in the $K$-band spectrum.
U Gem, on the other hand, has a lower mass transfer rate, high inclination, 
and weak CO absorption. Therefore the difference in observed CO absorption may be due to the
difference in binary inclination or mass transfer rate although the issue is not so clear
(see \S4).

\subsection{BPM 71213}

This binary consists of a non-magnetic WD plus an M dwarf 
with an orbital period of 4.33 hr.
The binary is non-interacting, presumably a pre-CV, and the secondary 
star shows chromospheric activity (Kawka et al., 2002).
These same authors determine the secondary star to be M2.5 V.

Our IR spectrum (Figure 6) shows typical absorption features of an M2-3 V star
including normal strength Mg I (2.11 \& 2.28 microns) and CO. 
Tappert et al. (2007) also observed BPM 71213 in the $K$-band and note
absorption features apparently weaker than expected for an M2-3 V star.
The higher S/N in the continuum may be the reason why we find
normal absorption depths for all the absorption features, including CO.
We observed BPM 71213 on both nights but see no difference in the $K$-band
spectrum. 
%No indication of the M star being chromospherically active
%was observed (nor expected in the $K$-band) for BPM 71213.

\subsection{QS Vir = EC 13471-1258}

EC 13471-1258 is suggested to be a hibernating CV albeit with an 
orbital period (P$_{\rm orb}$=3.6 hr)
outside of the usual CV period gap of $\sim$2 to $\sim$ 3 hr. 
If the mass transfer has stopped in this CV, it appears to have stopped quite early with
respect to its crossing the 2-3 hour period gap.
The binary contains a non-magnetic DA white dwarf and a completely normal, solar
abundance, chromospherically active M3.5-4 V secondary star (O'Donoghue et al., 2003).
The white dwarf is seen to have a rapid rotation possibly 
suggestive of evidence for past mass accretion.

Our infrared spectrum (Fig. 6) matches well with a template spectrum 
of a normal, single M3-4V star, and thus is
in good agreement with the O'Donoghue et al., (2003) result. 
Tappert et al. (2007) also observed this object in the $K$-band and note
absorption features weaker than expected for an M4V star.
As for BPM 71213, our higher S/N in the continuum may be the reason why we find
normal absorption depths for the absorption features, including CO.

The M4V secondary is noted to be very active, flaring often, and showing
the typical signs of chromospheric activity (O'Donoghue et al., 2003).
Our IR spectrum reveals no evidence for this behavior, however that is
completely expected as chromospheric activity does not manifest itself
in the higher energy hydrogen lines due to their formation deep in the stellar
photosphere (see \S 4.2, Howell et al., 2006).

The model fit to UV spectroscopy of the white dwarf (O'Donoghue et al., 2003)
implies a distance of only 48 pc and would make EC 13471-1258 the second closest 
CV, after WZ Sge (d=43 pc), if its CV nature is confirmed. 
The continuum flux in the V band is equally split between the
two stars while our $K$-band measurement will contain $<$3\% contamination from
the white dwarf. Assuming a 100\% contribution from a normal M4V star (M$_K$=
7.4) and convolving a K filter with our spectrum (giving K=11), we find a
distance to EC 13471-1258 of 52 pc in close agreement with the 
previous determination.

\subsection{SDSS0743, SDSS0757, SDSS0830}

These three possible pre-CV non-interacting binaries were discovered from the SDSS spectra taken of
red+blue objects (Silvestri et al. 2006a, 2007). An iterative fit to
the spectra with white dwarf and secondary star models produced
estimates for the temperatures of the white dwarfs and the spectral
types of the secondaries (Silvestri 2007). Additional followup spectra
and photometry revealed short orbital periods
of 3-4 hrs were likely (Silvestri et al. 2006b). We present our Keck $K$-band
spectra for all three in Figure 7.

SDSS0743 is a $\it{g}$=17.34 system with an M4V star determined
from the optical spectrum. Our $K$-band spectrum looks slightly earlier then M4V
although the Na I absorption is too strong for this type given the depth of the
CO absorption. The Na I strength would argue for a slightly later star, M5-M6V. 

SDSS0757 is a $\it{g}$=16.36 binary with
an M2V secondary as determined from optical spectroscopy. The $K$-band spectrum
shown in Figure 7 {\it may} show weak Br$\gamma$ emission but confirmation is
required. The weakness in the $K$-band of the MgI/AlI complex (2.11 microns) 
and the relative strength of CaI absorption at 2.26 microns argues for a slightly
later spectral type near M3-M4V.

SDSS0830 has a $\it{g}$ magnitude of 18.50
and an M0V secondary as determined from optical spectra. 
Our Keck spectra (Fig 7) is of low S/N but 
generally consistent with this type or slightly later
(M1-M2V) and appears to
contain blue continuum contamination from the white dwarf. 

%\subsection{SDSS1035}
%Star was not detected - remove from paper.

\subsection{SDSS0837}

This faint ($\it{g}$=19) LARP (Low Accretion Rate Polar) 
was identified from its unusual SDSS spectrum, which showed
cyclotron humps near 4500\AA\ and 6200\AA\ and TiO bands consistent
with an M5V secondary (Schmidt et al. 2005).
Their followup spectropolarimetry confirmed high polarization and determined
that the cyclotron humps were harmonics 4 and 3 in a field of 65 MG.
However, a polarization of the opposite sign was also apparent between
the harmonics, indicating that a second accretion pole with a different
field strength was also present. CCD photometry indicated an orbital
period of either 3.18 or 3.65 hr. Later XMM data and optical data
(Hilton et al. 2009) confirmed a very low accretion rate, consistent
with no accretion shock at the poles and determined the correct orbital
period to be 3.18 hr. While it is clear that LARPs do not have mass
transfer through the usual ballistic stream, it is likely they are still
accreting via a wind from the secondary that is channelled to the magnetic
poles (Howell 2008). Due to the very low accretion rates and especially to the low temperatures of the white dwarfs in LARPS (under
10,000K), it has been suggested that Roche-lobe overflow has not taken
place and these objects may be pre-Polars (Schmidt et al. 2005; Schwope
et al. 2009).

Our infrared spectrum is shown in Figure 8. While of low S/N, we note the presence of weak 
Na I and CO absorption. Ca I may be present, but the continuum noise level 
for this relatively faint target prevents us from a positive identification.
The secondary star in this LARP looks like a normal M4-5V 
star, similar to that observed in SDSS0743. 

\subsection{ST LMi}

The magnetic CV ST LMi has been the subject of many studies, most performed with optical or higher energy
observations and aimed at an understanding of the magnetic white dwarf or the accretion geometry and processes
(e.g., Cropper and Horne, 1994; Stockman and Schmidt 1996, Robertson et al., 2008). 
The white dwarf has a weak magnetic field for a polar, 11.5 MG, and the binary orbital period is short at 
1.9 hours, below the CV period gap. 

Figure 9 shows four consecutive spectral sets we obtained at 
Keck covering $\sim$30 min or about 0.25 of an
orbit. During this time, it is apparent that the strengths of the 
absorption features, assumed to arise from
the secondary star, vary, getting stronger toward the end of the 
spectral sequence (bottom of Figure 9).
Campbell et al. (2008) show that the $K$-band magnitude of ST LMi changes by 0.8
over an orbit causing absorption features to weaken as the n=4 cyclotron harmonic
dominates the continuum. During times of no, to weak cyclotron contamination, the
secondary star is observed to be M6V.
Emission lines of H and He are also apparent showing equivalent
width variations and profile shape changes, both
typical for this star as illustrated in Robertson et al. (2008). 

Changes in line strength, particularly Na I absorption, were noted previously for 
ST LMi by Howell et al
(2000) and attributed to star spots on the surface of the secondary star. 
We also see (as did Howell et al. 2000) that CO absorption bands are present, 
expected for a polar (Harrison et al. 2005) but 
they change strength with time. Changing spectral line profiles are common for polars
which present the observer with a
changing view of the cyclotron emission region during its orbit.

\subsection{RZ Leo}

This little studied, faint dwarf nova has an orbital period of 1.84 hr
(Mennickent and Tappert 2001). Previous work has been nearly exclusively
photometric and as such, we know little about the component stars in RZ Leo.
Mennickent \& Tappert (2001) present a RV solution with a K1 amplitude
of $\sim$80 km/sec. 

Our IR spectrum presented in Figure 10 covers a total time period of 
1 hour, or about one-half of the orbital period.
The emission lines due to Brackett$\gamma$ (2.16 microns) and He 2.056 microns
are of moderate strength and appear double-peaked. They are formed in the
accretion disk and suggest a high inclination, although no eclipses are observed.

Absorption features from the secondary star are weakly detected and consist of 
Na I, (strong) Ca I, and CO molecular absorption edges at 2.29, 2.32, and 2.35 microns. 
The third CO absorption band is blended with the Na I doublet at 2.35 microns, and the
fourth available CO absorption (2.38 microns) falls in a noisy continuum region
and we cannot be sure of its detection. While the absorption features are 
filled in by accretion flux, their relative depths are useful as spectral type
indicators. We find the secondary star in RZ Leo to be consistent with an 
M3-M4 V star, consistent with expectations (Smith \& Dhillon 1998).
RZ Leo is only the third dwarf nova (after IP Peg and SS Aur)
to show normal CO absorption in its infrared spectrum.  
WZ Sge shows CO as well, but in emission (Howell et al., 2004).

\subsection{GW Lib}

This short period CV has relatively rare superoutbursts and belongs to the 
Tremendous Outburst Amplitude Dwarf nova (TOAD)
class of CV (Howell, Szkody, \& Cannizzo 1995), similar to WZ Sge. 
Our summed IR spectrum covers nearly one complete orbital cycle
and is presented in Figure 10. We see a strong, narrow emission line (Br$\gamma$)
from the
accretion disk. The single-peaked nature suggests that the binary has a low
inclination as has been determined previously ($i$=11 degrees, Szkody et al., 2002; 
Thorstensen et al. 2002). A typical double-peaked 
accretion disk emission line profile in a system such as GW Lib 
(e.g., using WZ Sge as an example, 
Skidmore et al., 2000) would have a velocity width (FWZI) near 1200 km/sec for an inclination
of 90 degrees. At 11 degrees, this velocity width would be expected to be near 200 km/sec,
essentially at our spectral resolution but in agreement with the single-peaked line we see.

Our $K$-band spectrum has moderate S/N in the continuum but no 
explicit absorption features from the secondary star are detected. However it does
seem that the 
CO band (+ water vapor) continuum break, present in M8-9 stars and L type
brown dwarfs (see Kendall et al. 2004), may be marginally detected starting 
at 2.29 microns and extending redward. 
Given the strength of the white dwarf continuum (Szkody et al. have shown that the UV and
optical continua are dominated by the WD) and some accretion disk contribution 
(based on the presence of emission lines and the low inclination)
we are not surprised at the lack of secondary star absorption features.
If we are detecting the CO continuum break, our best fit donor star would be of
spectral type later then M9 V having a mass of less than 0.06 M${\odot}$.

Szkody et al., (2002) determined a mass of 0.8 M${\odot}$ for the pulsating white 
dwarf in GW Lib from fitting the HST spectrum with a dual
temperature white dwarf model.
Taking K1=40 km/sec, as found by Szkody et al. and
Thorstensen et al. (2002) from optical spectroscopy and our mass limit above, 
$q$ $\le$0.075 and the K2 velocity amplitude would be near 500 km/sec.

%\subsection{Tau 4}
%
%Tau 4 (RX J0502.8+1624) has been previously studied by
%Littlefair et al. (2005) and Howell et al. (2008). 
%The probably short period binary is a magnetic CV
%with cyclotron humps clearly detected in the $K$-band by Howell et al. 
%The $K$-band spectra also show H and He I emission and a short-term absorption phase
%associated with a gas stream crossing. 
%
%We present a single mean Tau 4 spectrum 
%in Figure 4 which shows no evidence of any absorption features nor the
%CO band continuum break. This single spectrum will serve our purposes 
%and we refer the reader to Howell et
%al. (2008) who present time-series IR spectroscopy of Tau 4.

\section{Discussion}

Our sample of pre- and present day CVs, both with magnetic and non-magnetic white dwarfs,
was obtained in order to provide a more coherent picture of the presence
or absence of CO absorption in the IR spectra of these stars. Harrison et al. (2004,2005)
set the stage by showing the apparent presence of normal CO absorption in polars and the
apparent lack of or weak CO absorption (but the presence of O) in dwarf novae. 
Taking this information, it has been suggested that the secondary stars in dwarf novae
are highly evolved CNO burners, while those in polars are younger main sequence
stars of near solar abundance. This scenario requires a different evolution for pre-CVs
which become dwarf novae as compared to pre-CVs that become polars - the latter group
being essentially unknown (Liebert et al. 2005). 
The results presented here appear to add further confusion to the issue of the 
evolutionary path of CVs. We find that the polar V1309 Ori has extremely
weak CO features, in contrast to the result for all other polars (Harrison
et al. 2005), while the secondary in the mainstream non-magnetic CV, SS Aur, 
appears to have normal CO features. Both of these objects have a few 
characteristics that set them apart from other members of their respective 
classes, but they certainly do not show any behavior that is not seen in
other members of their classes. We confirm the suggestion by Dhillon et 
al. (2002) that there is no evidence for enhanced 13CO absorption in 
V471 Tau. Thus, 
none of the so-called pre-CVs presented here, or elsewhere, show any sign of 
abundance peculiarities. 

The objects observed in this study are listed in Table 2 and 
our former $K$-band studies in the literature are listed in Table 3. 
Taken together, these observations seem to cloud the issue of CO absorption 
as we find that all the
pre-CVs, which all have non-magnetic white dwarfs and are thought to 
evolve into dwarf
novae, show apparent normal abundances including CO absorption. 
We also find that the supposedly
hibernating CV, QS Vir, shows normal CO absorption as well. In addition, one polar
(V1309 Ori), does not show CO absorption but has the somewhat non-typical 
polar properties of a very long orbital period and a subgiant secondary.
There is no single theory to reconcile these results that doesn't strain 
credulity. If magnetic and non-magnetic CV evolution is identical, then we must 
ignore the growing evidence from UV and IR observations that there appear to be 
abundance issues in non-magnetic CV systems. This requires us to find a method to
preferentially obscure the CO features in the secondary star of disk-accreting
systems (CO emission from the accretion disk?), 
and (simultaneously) excite nitrogen emission over that of carbon. Or we
must come up with a method that strongly enhances N, and depletes C, between the
pre-CV phase and contact, that {\it only} occurs in disk-accreting CVs. This also 
relegates some subset of CVs, including U Gem, WZ Sge, VY Aqr, V1309 Ori, 
AE Aqr, GK Per, CH UMa, and perhaps even SS Cyg (Robinson \& Bitner 2009) and RU 
Peg, to the "oddball" bin. 
%The present observations show that 
%the ratio of normal to weak to absent CO absorption among well observed 
%pre-CVs and CVs (magnetic and non-magnetic), 22 in total so far, 
%is 7:10:5 respectively.

The idea of actual low C abundance would likely mean 
that some of the secondary stars follow a different evolution to become a CV with a
seemingly high dependence on their white dwarf's magnetic properties. 
Another possibility is that some of the present day CVs we observe are simply systems which 
contain older secondary stars, stars that are highly evolved nearing or just past their
main sequence lifetime. If this latter idea is true, then it implies that essentially all
polars are younger systems and any containing evolved secondaries are essentially unknown
or observationally selected against.
Additionally, the present sample of pre-CVs seems not to contain any evolved stars
soon to be CV donors.
We could try to accept that pre-CVs secondaries go
through an extremely rapid evolution including CNO burning between the present epoch and
their becoming a dwarf novae or that all the present day pre-CVs will not actually end up
as dwarf novae.  But both of these scenarios seem highly ad hoc.
It may be that the entire
binary undergoes a different evolutionary path depending on the formation of a magnetic
or non-magnetic WD (Tout et al., 2008) with some non-magnetic systems taking longer until
contact allowing some donors to evolve. 

Another scenario for low C abundance may be that the
secondary star collects CNO processed material from the white dwarf during DN outbursts
and this material covers and hides the normal abundance photosphere providing us with
a view showing an apparent lack of CO. This idea has many flaws, not the least of which 
are the
amazingly small solid angle the secondary subtends in which to receive most of the 
material,
the problem that this material would have to spread to cover the
entire secondary star surface, and the fact that a convective atmosphere will mix the
material into the star's 
interior very fast, a few tens of minutes (Kawaler 2008, priv. comm).

Exploring another option, Howell et al. (2004) noted that WZ Sge, a non-magnetic
CV, shows CO {\it emission} in its $K$-band spectrum. The most likely formation site 
for the CO
emission was in the outer cool dense regions of the accretion disk. These regions are
dense and cool enough to shield the CO molecules from the UV emission able to 
disassociate them. Furthermore, Howell
et al. (2008) found that the outer accretion disk regions in WZ Sge were indeed very
cool, cool enough for dust to form. The common property linking many of the CVs which show no or weak
CO absorption is the fact that they have accretion disks -- most of those without accretions disks
show normal CO absorption.
QS Vir may be the most interesting. It is supposed to be a hibernating non-magnetic CV, 
thus it was a DN with an accretion disk and will be again,
yet it has a normal looking secondary star including CO absorption. 

Can CO emission from the accretion disk fill in the secondary star 
CO absorption and account for the weak or missing lines? 
To see if this might work, we choose to model the dwarf nova
U Gem which is noted to have very weak to no CO absorption present in its $K$-band spectrum
(Harrison et al., 2005b). 
We produced an artificial CO emission spectrum, and using the fact that the   
Br$\gamma$ disk emission line in U Gem
has a FWHM velocity of $\sim$1750 km/s, we broadened our model CO emission
by 1500 km/s so as to approximate a more ``distant" emission region (i.e., the outer disk). 
We next took an M4V template spectrum (the spectral type of the secondary star in U Gem)
and summed it with our CO emission model.
The M4V template spectrum we used is
shown at the bottom of the plot. It has a velocity broadening of 90 km/s applied to get
it to look more or less like the observed secondary star features present 
in the U Gem spectrum.
 
We normalized the first overtone of the CO emission spectrum to exactly match
the absolute depth of the CO overtone absorption band 
in the template spectrum to make a best attempt to "fill-in"
the CO absorption spectrum. We then
replaced the red end (redward of 2.28 microns) of the U Gem $K$-band spectrum
(orbital phase 0.20) with our summed 
model scaled to match the observed continuum at 2.28 microns (Figure 11).
While we did not produce a full disk model spectrum with added CO emission,
this exercise was intended to illustrate whether CO emission could fill-in or hide
the expected M star molecular absorption. The spectrum labeled ``100\%" in Fig. 11
is our initial model as just described while the one labeled 
``50\%" is the same process but with the CO overtone emission at only 
50\% of its initial strength.  
We note that in the 100\% case, the narrow CO absorption features create
troughs at the emission line peaks
(similar to the HI lines in GK Per and SS Cyg as shown in Fig. 10 of Harrison et al., 
2009). Even in the ``50\%" spectrum is choppy with emission humps on both sides of 
each of the narrower CO absorption features. 
However, when the S/N of the red continuum is poor, this choppiness
may not be noticeable as in, for example, the spectrum presented here for RZ Leo and other
fainter DN with $K$-band spectra in the literature.

If CO emission from the accretion disk is the cause of the weak or absent secondary star 
CO absorption then why do not the
systems where the $K$-band spectra are completely dominated by disk emission
show CO emission? Only WZ Sge has shown clear CO emission and we noted earlier that this
same star is also the only CV which is carbon rich. 
The CV systems where the disk flux dominates the
$K$-band (i.e., no secondary star seen) are ones with high mass transfer rates and 
hot accretion disks, perhaps too hot to contain cool regions able to form CO. 
Or they are low inclination systems
where the hot inner disk and white dwarf continuum light 
outshine the remaining disk (including any CO emission present). 

There are some non-magnetic CVs which show weak or 
normal CO absorption strengths yet also have accretion disks.
We could try to make arguments such as in RZ Leo we see
only weak CO absorption perhaps related to its high inclination ($i$=65 degrees) 
or its low mass transfer rate.
SS Aur, on the other hand, has a moderate inclination ($i$=38 degrees) 
but contains a hot (30,000K) WD. Perhaps this hot star provides sufficient UV flux
so as to not allow CO to form in the disk. Maybe the carbon abundance is just not high
enough to produce CO emission in the accretion disk? 
Do accretion disks {\it know} how to make just enough CO
emission to fill-in the secondary stars CO absorption in the ``no" cases and just too little in the weak
cases? Why is WZ Sge so far the only disk system to show CO emission? Does the lack of CO absorption
in a DN remain constant over the orbit and over time?  All good questions without good answers.

Based on weak CO absorption, parallaxes, and other abundance
anomalies, Harrison et al. (2004a) 
have argued that the CVs SS Cyg, RU Peg AE Aqr, and GK Per contain subgiant
secondary stars. We show in this paper that the long-period polar V1309 Ori is likely to contain a
subgiant secondary as well.
Binary evolution models of CVs (e.g., Howell et al., 2001) show that systems in the range of 
about 3 to 5 hours have bloated secondary stars, that is their secondaries are too large for their mass
compared to main sequence models. This fact may make some secondaries appear to be subgiants 
but cannot alone explain the weakness of CO absorption, 13CO, or other anomalies in the
spectra.
Beuermann et al. (1998) explored evolutionary scenarios in which the secondary stars had undergone
nuclear evolution prior to mass transfer leading to the suggestion that many long period CVs should have
evolved donor stars.
Barraffe and Kolb (2000) have shown that some observational properties of CVs above the period gap can
be explained if evolutionary models include secondary stars which have some H depletion in their core,
including secondaries which are substantially evolved off the main sequence (i.e., subgiants).
Detailed abundance studies of CV secondaries, while a difficult observational task, is needed to place
any of these arguments on a firm foundation. 

The reality is that
we do not know the exact mechanism by which CO emission could
be produced in the accretion disks or in what fraction of disks
it might be produced.
Some CVs show evidence in both the UV (high N/C ratios) and the $K$-band (weak or absent CO)
to suggest that the secondary stars are highly evolved.
Our sample of 24 disk systems discussed here 
does not provide any consistent clues to settle the discussion.
Perhaps a variety of mechanisms are operating simultaneously within the binary systems.
To date, there does not seem to be a
single answer which can account for all the observations regarding the weak or absent CO absorption 
in the secondary stars of cataclysmic variables.

%We might now postulate that weak or a complete lack of CO absorption observed in CVs
%may be related to the presence of CO emission formed in the outer accretion disk regions. 
%Other than in WZ Sge, there is no direct evidence for this statement, only circumstantial 
%evidence based on the fact that when donor star absorption lines are observed in CVs which have 
%accretion disks, weak or no CO absorption occurs. However, not all CVs 
%follow this rule.
%Observational evidence in 25 or so CVs to date points to abundance anomalies and suggests that donor
%stars may be highly evolved with respect to typical 
%main sequence stars and some have been shown to likely 
%be subgiants.
%Theoretical models have been constructed to attempt to match the present-day CV population as well as
%the observational constraints with evolved donors playing a part. Quantitative abundance analysis
%for any donor star in a CV has yet to be performed as it is a difficult observational task.
%However, such work would shed light on the evolutionary state of CV donor stars and greatly help
%to solve the issues discussed herein.

\acknowledgments
We wish to thank the Keck II OAs Heather, Chuck, Terry, and Jason for their help
with our observations and Jim Lyke for his help in the instrument setup.
SH thanks the ESO DFGR for Financial support and Marina Orio and the University of
Padova Observatory, for their hospitality and the use of a 16th century
Specola tower office. We thank the anonymous referee for their detailed and careful reading
of the manuscript.

Data presented were obtained at the W. M. Keck Observatory, which is operated
as a scientific partnership among the California Institute of Technology, the
University of California, and the National Aeronautics and Space Administration. The
Observatory was made possible by the generous financial support of the W. M. Keck
Foundation. The authors wish to recognize and acknowledge the very significant
cultural role and reverence that the summit of Mauna Kea has always had within the
indigenous Hawaiian community. We are most fortunate to have the opportunity to
conduct observations from this mountain.

%+++++++++++++++++++++++
%Table 1 - Observing Log
\begin{deluxetable}{cccc}
\tablenum{1}
\tablewidth{6.2in}
\tablecaption{Observing Log}
\tablehead{
  \colhead{Star}
& \colhead{UT Date}
& \colhead{UT Start}
 & \colhead{Total Exp. Time}
}
\startdata
\hline
V1309 Ori & 4 Mar 2007 & 5:32 & 16 min   \\ 
V471 Tau & 4 Mar 2007 & 5:22 & 12 sec   \\
QS Vir & 4 Mar 2007 & 14:48 & 4 min  \\
BPM 71213 & 4 Mar 2007 & 5:09 & 4 min \\
BT Mon & 4 Mar 2007 & 6:22 & 40 min  \\
GW Lib & 4 Mar 2007 & 15:23 & 64 min  \\
SS Aur & 5 Mar 2007 & 7:09 & 32 min  \\
RZ Leo & 5 Mar 2007 & 11:58 & 16 min  \\
ST LMi & 4 Mar 2007 & 13:11 & 64 min  \\
SDSS0837 & 4 Mar 2007 & 10:43 & 8 min  \\
SDSS0830 & 4 Mar 2007 & 9:44 & 8 min  \\
SDSS0757 & 4 Mar 2007 & 8:49 & 8 min  \\
SDSS0743 & 4 MAr 2007 & 7:56 & 16 min  \\ 
HS1136 & 4 Mar 2007 & 12;30 & 4 min  \\
\hline
\enddata
%\tablenotetext{a}{Four, 16 min exposures were obtained.}
\end{deluxetable}

%Table 2 - CO Status
\begin{deluxetable}{cccccl}
\tablenum{2}
\tablewidth{6.2in}
\tablecaption{Status of CO absorption-This Sample}
\tablehead{
  \colhead{Star}
& \colhead{Type}
& \colhead{P$_{orb}$ (hr)}
 & \colhead{Magnetic?}
 & \colhead{CO Ab.$^a$?}
 & \colhead{Notes}
}
\startdata
\hline
HS1136 & pre-CV & 20.1 & N & N & WD dominates IR flux \\
V471 Tau & pre-CV & 12.5 & N & Y & \\
BT Mon & DN & 7.99 & N & N & Ecl., disk dominates IR flux \\
V1309 Ori & polar & 7.98 & Y & N & Ecl. \\
SDSS0743 & pre-CV & 4.6 & N & Y & \\
SS Aur & DN & 4.38 & N & Y & \\
BPM 71213 & pre-CV & 4.33 & N & Y & \\
QS Vir & hibernating CV & 3.6 & N & Y & =EC 13471 \\
SDSS0757 & pre-CV & 3.5 & N & Y & \\
SDSS0837 & LARP & 3.18 &Y & Y & pre-polar? \\
SDSS0830 & pre-CV & 2.9 & N & Y & \\
ST LMi & polar & 1.91 & Y & Y & cyclotron cont. \\
RZ Leo & DN & 1.84 & N & W & \\
GW Lib & DN/TOAD &  1.33 & N & N? & \\
\hline
\enddata
\tablenotetext{a}{Y=appears normal for spectral type; W=appears weaker than normal for spectral type; N=not
present; N?=maybe present, too low S/N to be certain.}
\end{deluxetable}
%+++++++++++++++++++++++

\scriptsize
%Table 3 - CO Status
\begin{deluxetable}{cccccl}
\tablenum{3}
\tablewidth{6.2in}
\tablecaption{Status of CO Absorption-Previous Sample$^a$}
\tablehead{
  \colhead{Star}
& \colhead{Type}
& \colhead{P$_{orb}$ (hr)}
 & \colhead{Magnetic?}
 & \colhead{CO Ab.$^b$?}
 & \colhead{Ref$^c$}
}
\startdata
\hline
AE Aqr & DQ & 9.86 & Y & W & 7 \\
SY Cnc & DN & 9.12 & N & N & 5 \\
RU Peg & DN & 8.99 & N & W & 5 \\
CH UMa & DN & 8.23 & N & W & 5 \\
MU Cen & DN & 8.21 & N & W & 5 \\
AC Cnc & NL & 7.21 & N & W & 5 \\
EM Cyg & DN & 6.98 & N & Y & 5 \\
V426 Oph & DN & 6.85 & N & Y & 5 \\
SS Cyg & DN & 6.60 & N & Y & 5 \\
AH Her & Z Cam & 6.20 & N & W & 5 \\
BV Pup & DN & 6.35 & N & W & 5 \\
EX Dra & DN & 5.04 & N & Y & 4 \\ 
SDSS1553+55 & polar & 4.39 & Y & Y & 3 \\
TW Vir & DN & 4.38 & N & N & 4 \\
SS Aur & DN & 4.29 & N & W & 4 \\
U Gem & DN & 4.25 & N & W & 4 \\
UU Aql & DN & 3.92 & N & N & 4 \\
IP Peg & DN & 3.80 & N & Y & 4 \\
RR Pic & CN & 3.48 & N & W & 4 \\
AM Her & polar & 3.09 & Y & Y & 4 \\
AR UMa & polar & 1.93 & Y & Y & 3 \\
ST LMi & polar & 1.90 & Y & Y & 3 \\
MR Ser & polar & 1.89 & Y & Y & 4 \\ 
VV Pup & polar & 1.67 & Y & Y & 2,3 \\
VY Aqr & TOAD & 1.51 & N & N & 1 \\
EI Psc & DN? & 1.07 & N & N & 1 \\
WZ Sge & TOAD & 1.35 & N & E & 6 \\
\hline
\enddata
\tablenotetext{a}{Six previous $K$-band spectra from the references had too low of S/N to provide a CO
comment. Three low field polars were not included as they have their $K$-band continuum dominated by
cyclotron emission. AC Cnc and EX Dra eclipse.}
\tablenotetext{b}{Y=appears normal for spectral type; W=appears weaker than normal for spectral type; N=not
present; E=emission.}
\tablenotetext{c}{1=Harrison et al., 2009, 2=Howell et al. 2006a, 3=Harrison et al., 2005a, 4=Harrison et al.,
2005b, 5=Harrison et al., 2004., 6=Howell et al., 2004, 7=Harrison et al., 2007}  
\end{deluxetable}
\normalsize
%+++++++++++++++++++++++

\pagebreak

% ----------- figures -------

\begin{figure}  % figure 1  
\epsscale{0.7}
%\plotone{NEWFIG0.ps}
\plotone{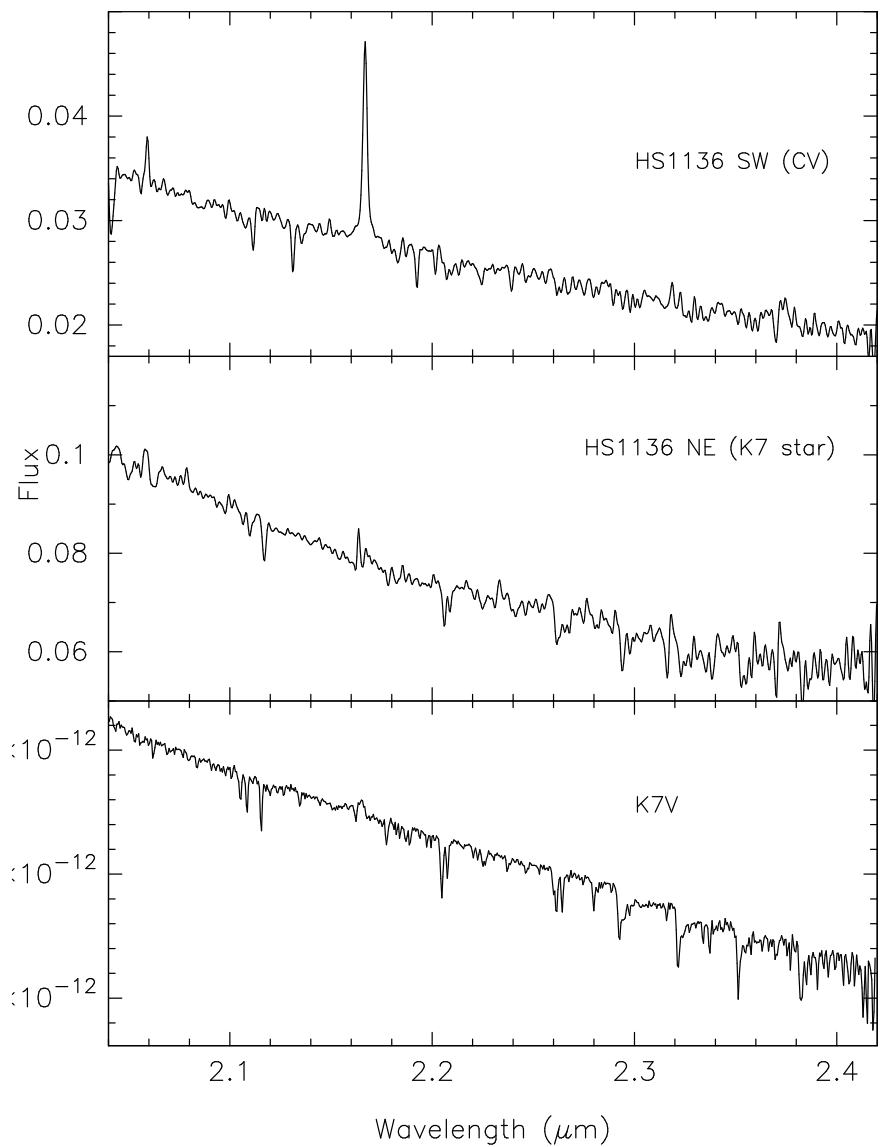}
\caption{Our infrared $K$-band spectrum 
of the pre-CV HS1136 (top) and its spatially nearby, 
common proper motion companion K7V star (middle).
%Residual Br$\gamma$ emission in this star 
%is due to slit contamination by the pre-CV star only 1.3 arcsec away.
%HS1136's $K$-band light is dominated by the continuum flux from the hot 
%(70,000K) white dwarf. Thus, the M2.5V dwarf companion is not visible in optical thru
%near-IR light except in the optical Balmer emission lines due to irradiation. 
The bottom panel shows a template single K7V star for comparison.
In all the Figures, the Y-axis is relative flux either presented as normalized for
our program stars and some templates (set to 1.0 at 2.19 microns) 
or presented in units of ergs/cm$^2$/sec/\AA~for most of the template standards.
}
\end{figure}

% ---------------------------------

\begin{figure}  % figure 2 
\epsscale{1.0}
%\plotone{cfig2.ps}
\plotone{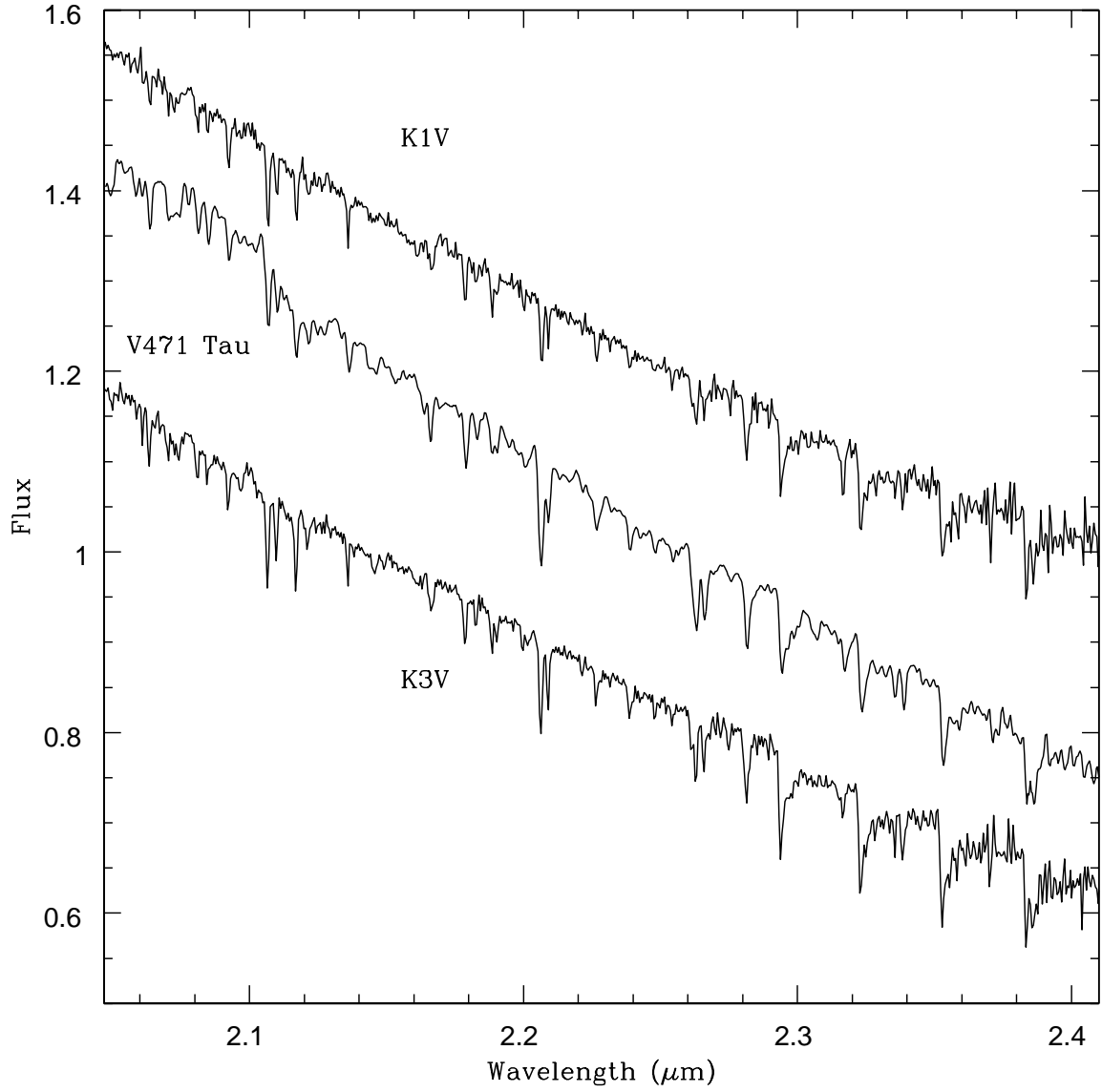}
\caption{Infrared $K$-band spectrum of the canonical pre-CV V471 Tau compared with 
single star templates for K1V and K3V stars. 
}
%\label{longlc}
\end{figure}
% ---------------------------------

\begin{figure}  % figure 3 
\epsscale{0.7}
%\plotone{NEWFIG1.ps}
\plotone{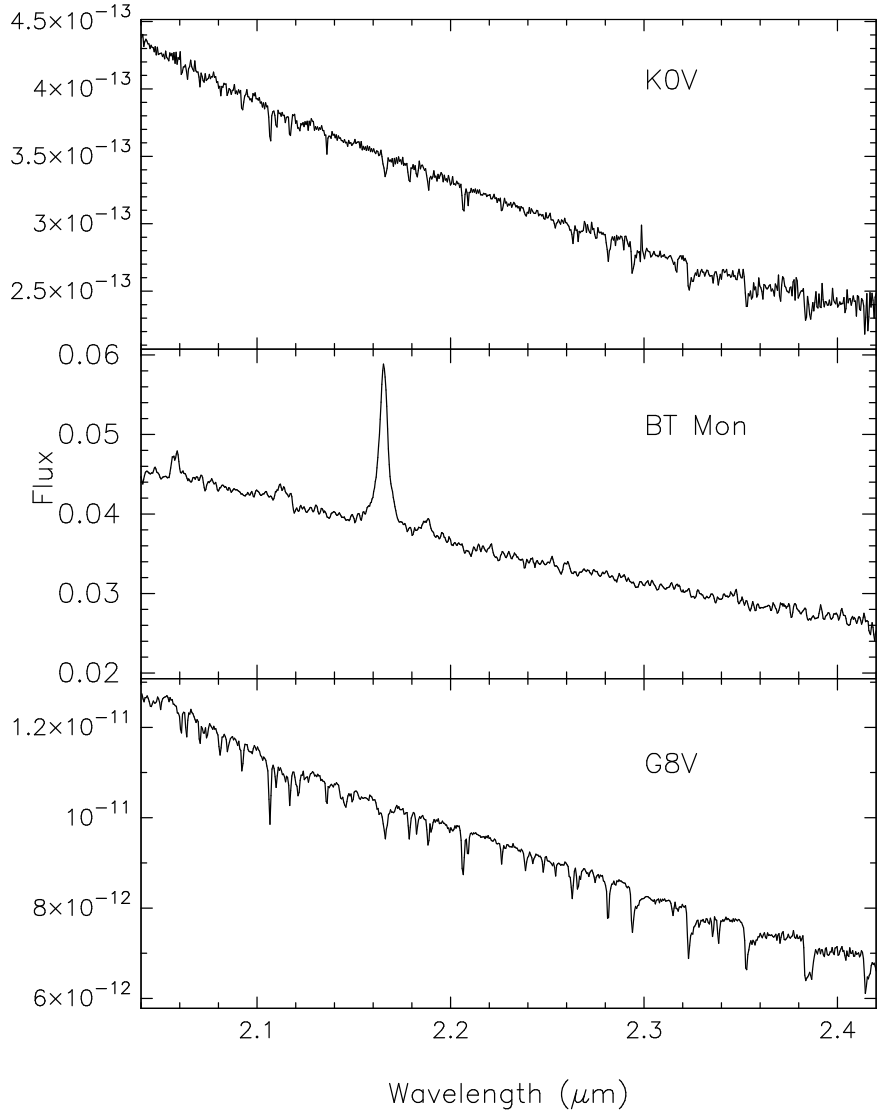}
\caption{Infrared $K$-band spectrum of the old nova BT Mon 
(containing a G8V secondary star). We show single star templates for K0V and G8V
in the figure.
% but note that BT
%Mon's $K$-band light is dominated by flux from the $i$=11 degree bright accretion disk.
%No sign of the secondary star is seen at this S/N level for the continuum. 
}
%\label{longlc}
\end{figure}
% ---------------------------------

\begin{figure}  % figure 4  
\epsscale{0.85}
%\plotone{cfig1a.ps}
\plotone{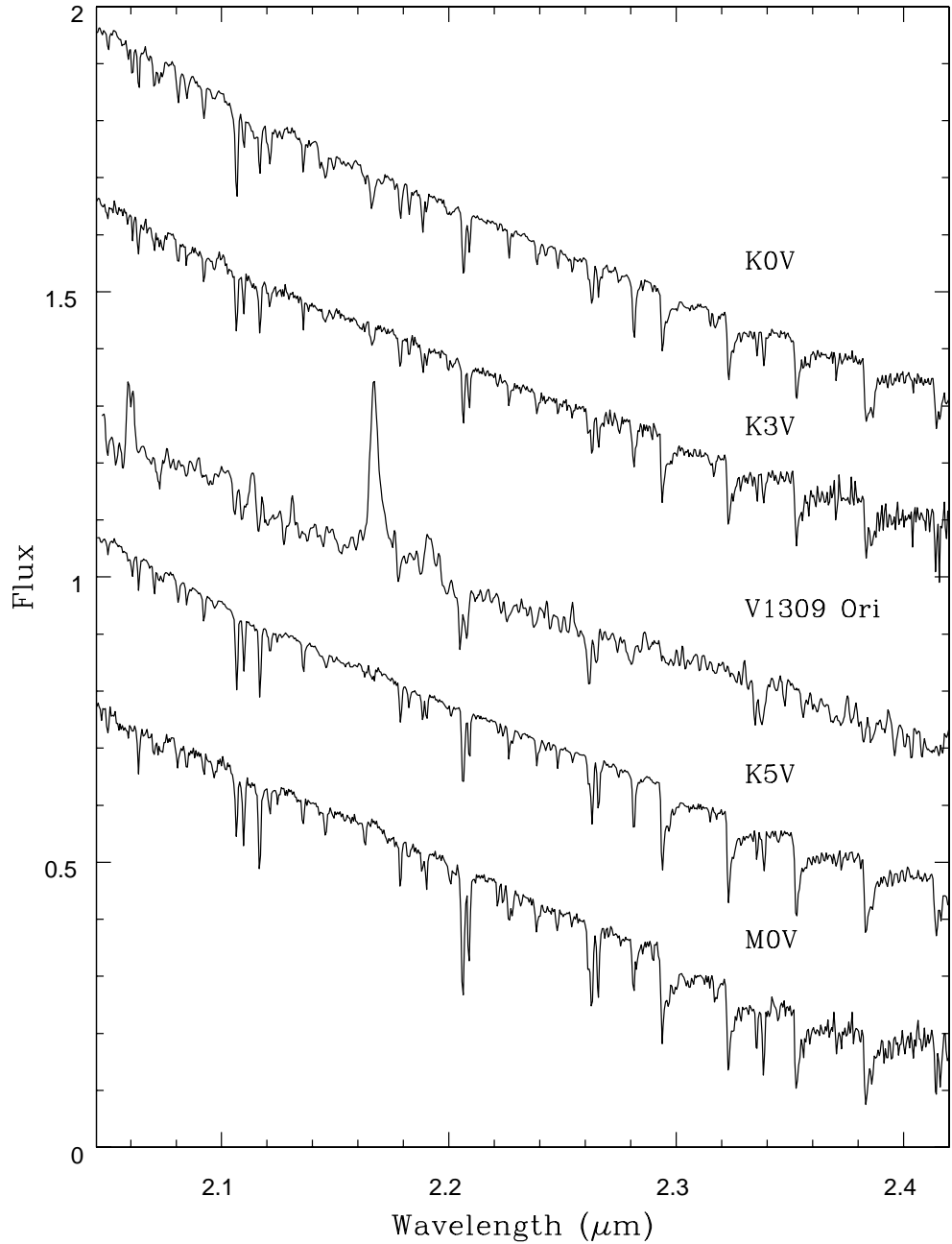}
\caption{Comparison of V1309 Ori $K$-band spectrum with single star template spectra
covering M0V to K0V. The line ratios for NaI to CaI are consistent with a secondary star
of spectral type near K7. Note the complete absence of CO absorption in V1309
Ori.
}
%\label{longlc}
\end{figure}

% ---------------------------------

\begin{figure}  % figure 5
\epsscale{0.7}
%\plotone{SSAurcomp.ps}
\plotone{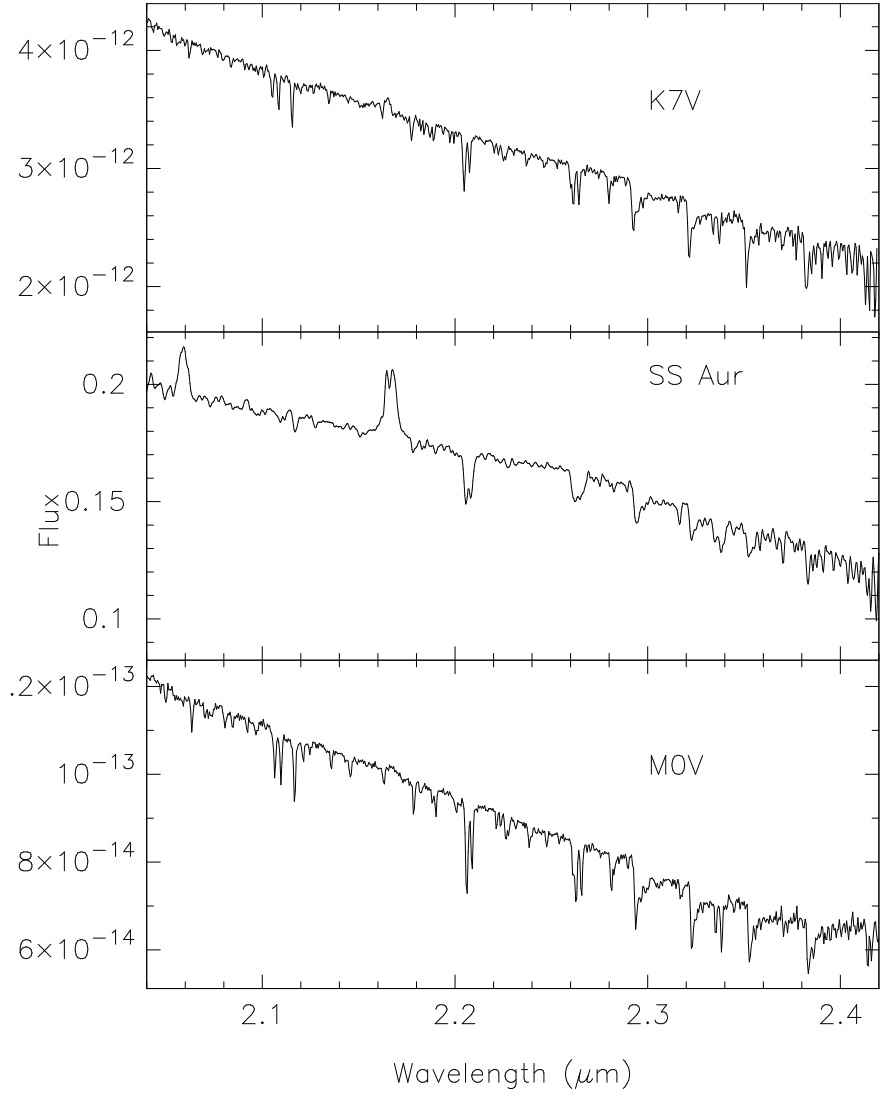}
\caption{Infrared $K$-band spectrum of the dwarf nova SS Aur. 
%SS Aur contains a 30,000K white dwarf and an M secondary star very close to M4V.
%We show single template $K$-band spectra for M2V and M4V stars in the figure.
%SS Aur shows CO and other absorption features with ratios similar to early M
%stars although the Na I and Ca I line strengths are a bit too strong.
The double-peaked Br$\gamma$ line is due to the
accretion disk in this binary with an inclination of 38 degrees.
}
%\label{longlc}
\end{figure}

% ---------------------------------

\begin{figure}  % figure 6
\epsscale{0.7}
%\plotone{NEWFIG3.ps}
\plotone{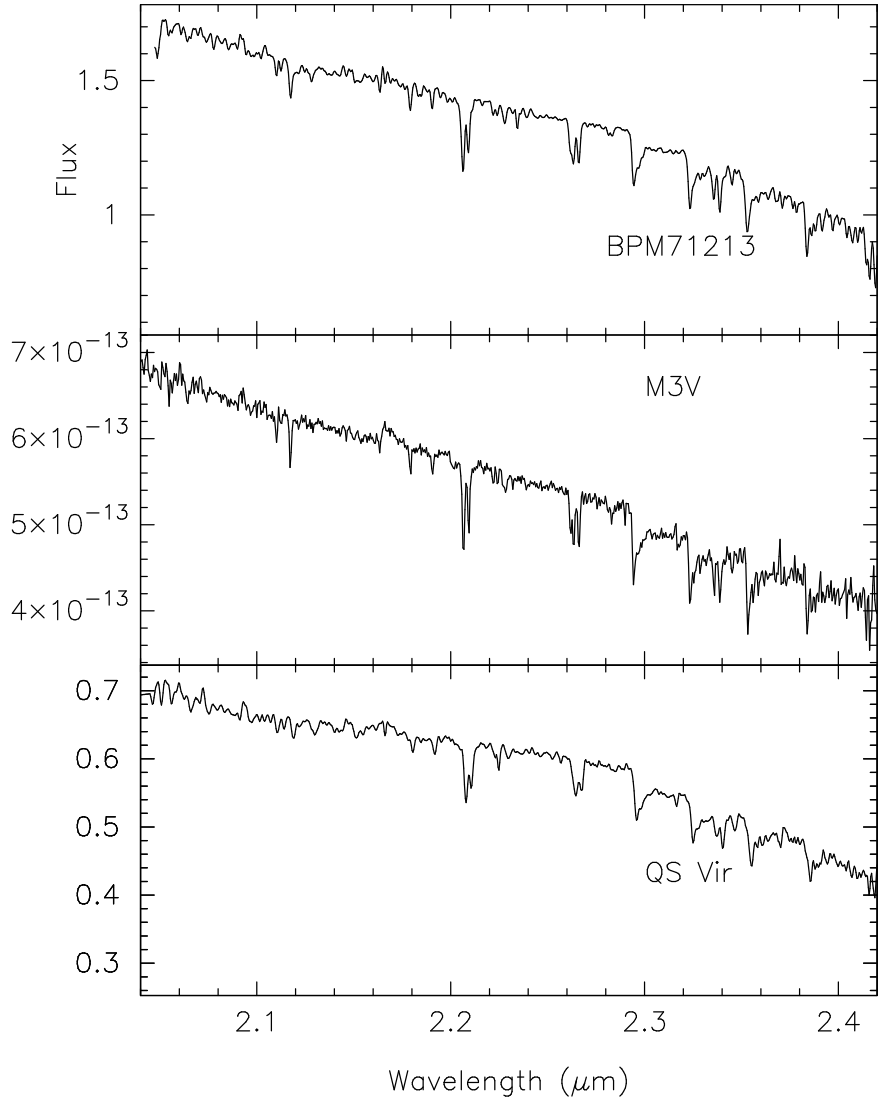}
\caption{Infrared $K$-band spectra of the pre-CV BPM71213 and the hibernating CV QS
Vir. Both systems have secondary stars near M2V-M4V and we show a single template
spectrum of a M3V for comparison. Note that both stars show normal CO and other
absorption bands.
}
%\label{longlc}
\end{figure}

% ---------------------------------

\begin{figure}  % figure 7
\epsscale{0.7}
%\plotone{NEWFIG4.ps}
\plotone{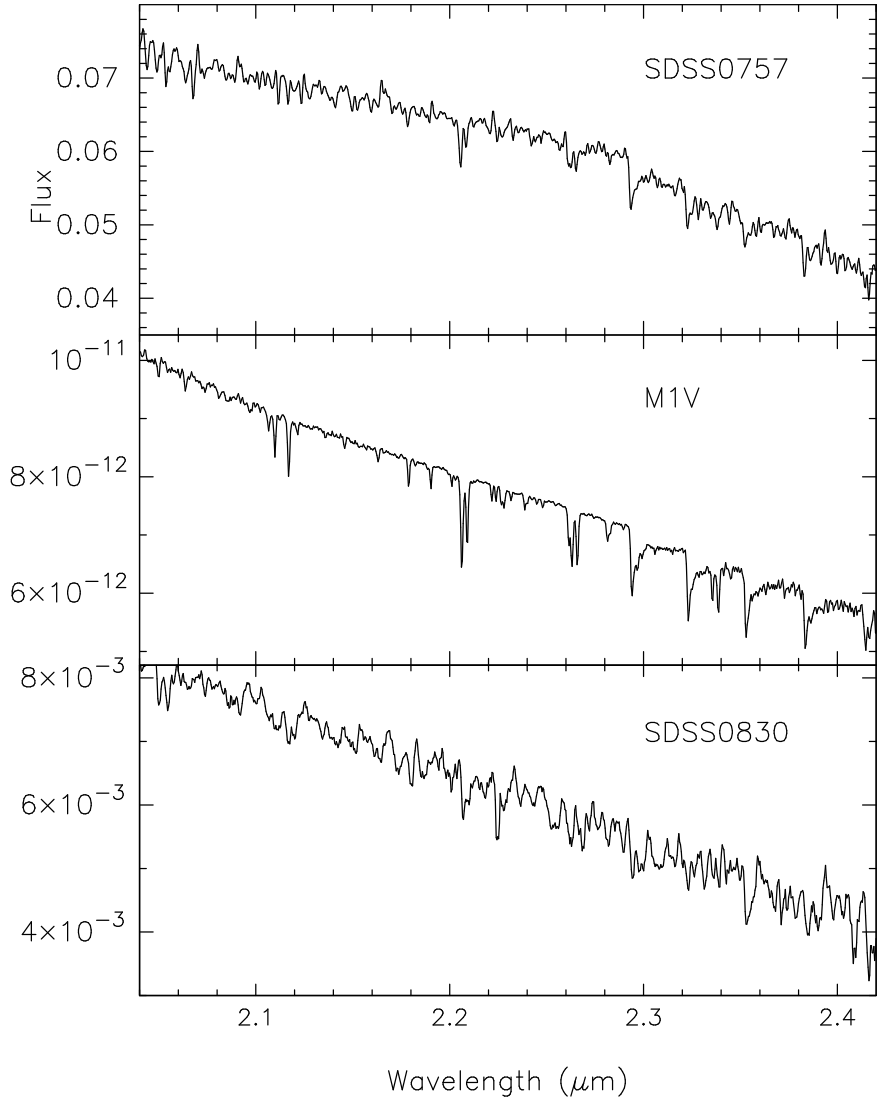}
\caption{Infrared $K$-band spectra of the pre-CVs SDSS0757 and SDSS0830.
Both systems have secondary stars near M0V-M2V and we show a single template
spectrum of a M1V for comparison. Note that both stars show normal CO and other
absorption bands.
}
%\label{longlc}
\end{figure}

% ---------------------------------

\begin{figure}  % figure 8
\epsscale{0.7}
%\plotone{NEWFIG5.ps}
\plotone{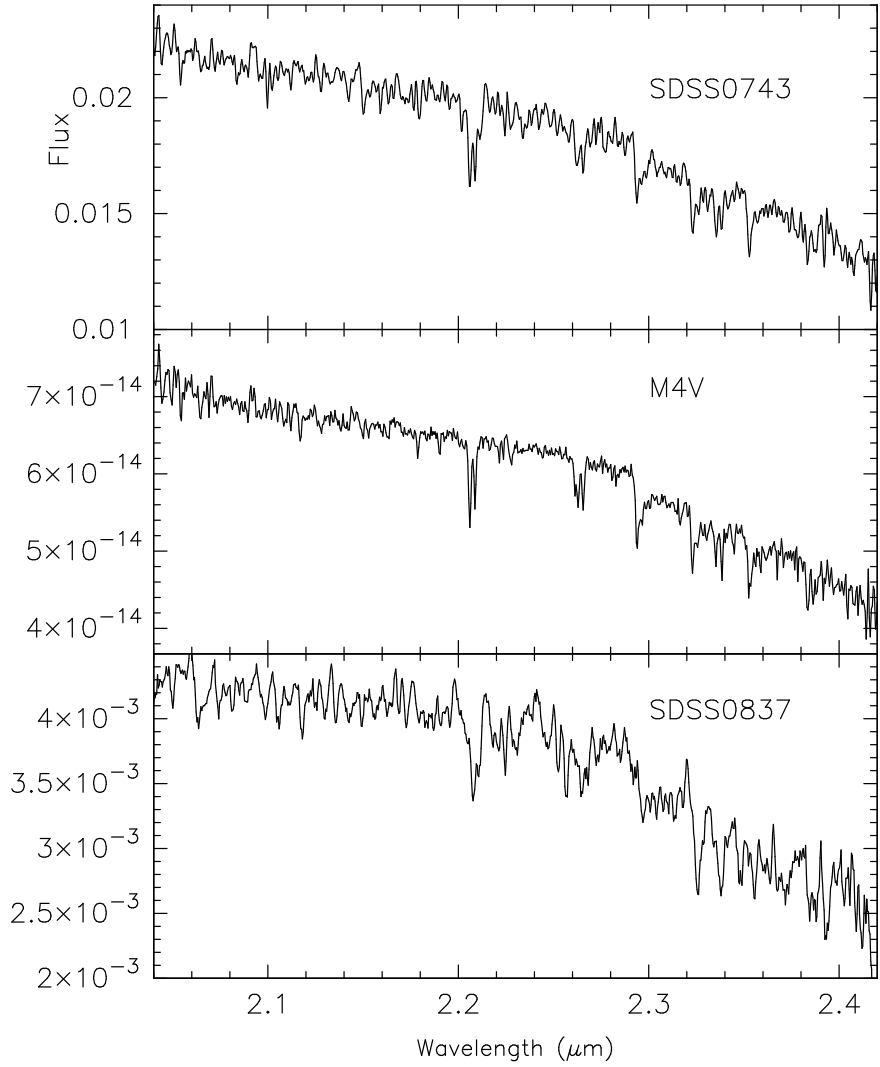}
\caption{Infrared $K$-band spectra of the pre-CV SDSS0743  and the LARP (pre-polar) 
SDSS0837.
Both systems have secondary stars near M4V and we show a single template
spectrum of a M4V for comparison. 
%Note that both stars show normal CO and other
%absorption bands although is it difficult to tell exactly for the low S/N spectrum of
%SDSS0837.
}
%\label{longlc}
\end{figure}

% ---------------------------------

\begin{figure}  % figure 9 
\epsscale{1.0}
%\plotone{cfig5.ps}
\plotone{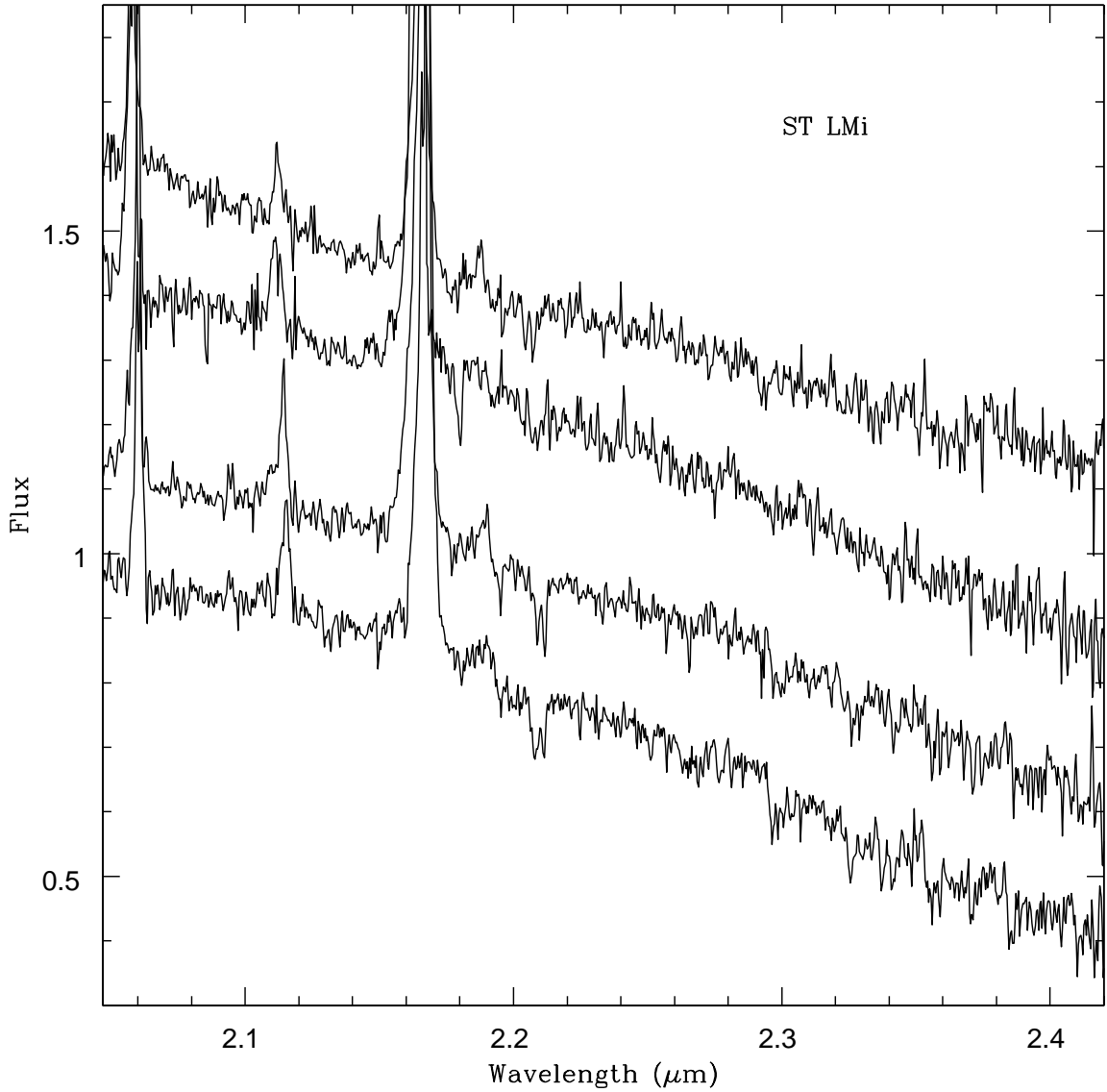}
\caption{Four infrared $K$-band spectra of ST LMi covering a short segment (4X16 minutes) of its full orbital period (1.09
hr). We note the short-term changes in the absorption lines and bands as evidenced by the Na doublet at 2.20/2.335 microns
and the CO bands at 2.29 microns. 
%Emission line changes over the orbit are common for ST LMi due to the coming and going of
%cyclotron emission. The associated highly modulated cyclotron continuum in the $K$-band is likely to be the cause of the
%apparent absorption line changes.
}
%\label{longlc}
\end{figure}

% ---------------------------------

\begin{figure}  % figure 10
\epsscale{0.7}
%\plotone{NEWFIG6.ps}
\plotone{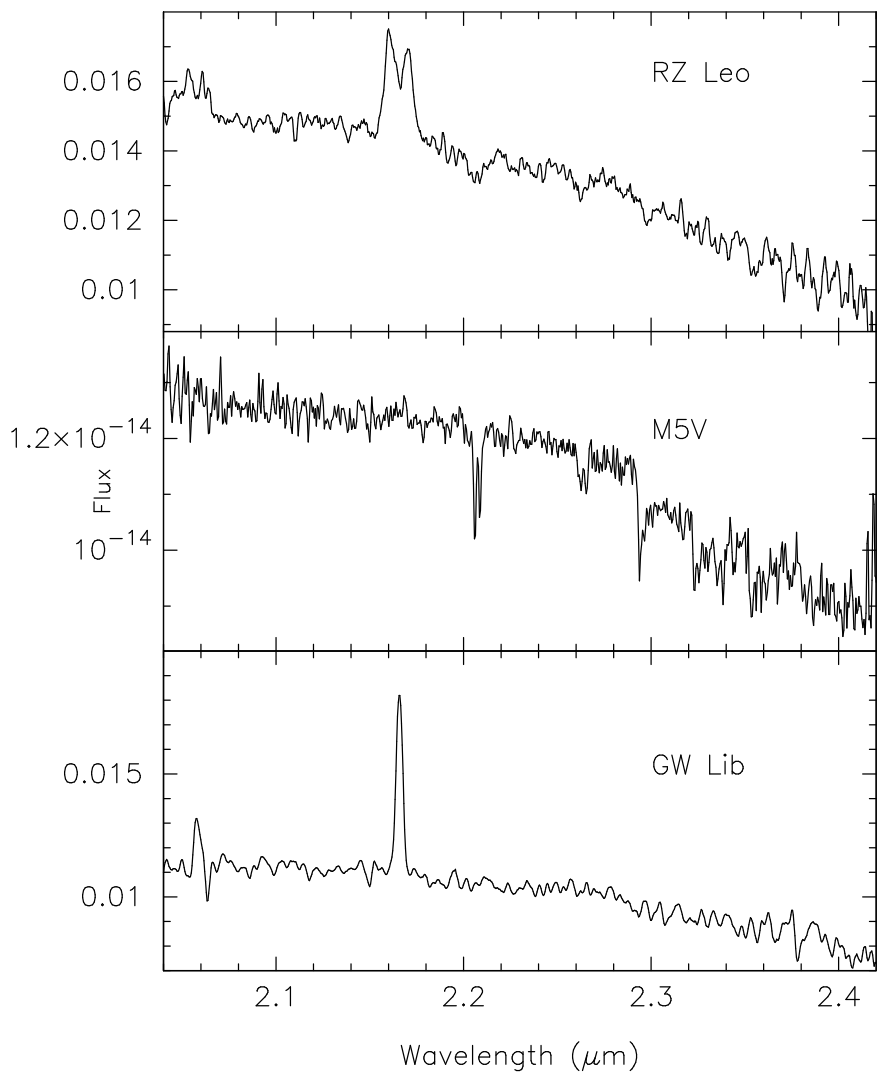}
\caption{Infrared $K$-band spectra of the TOAD RZ Leo believed to have a M5V-like
companion. Weak CO and other absorption
lines are seen and we show a single star template spectrum of a M5V for comparison.
%The absorption lines are either weak or filled in by accretion disk light or other
%flux. 
The bottom spectrum is of the TOAD GW Lib showing string emission in
Br$\gamma$. We see no direct evidence of the secondary star, expected to be near M9V
or a brown dwarf-like star, but do note the steam band/CO continuum break near 2.28
microns and redward.
}
\end{figure}

% ---------------------------------

\begin{figure}  % figure  11
\epsscale{1.0}
%\plotone{UGemWithCO.ps}
\plotone{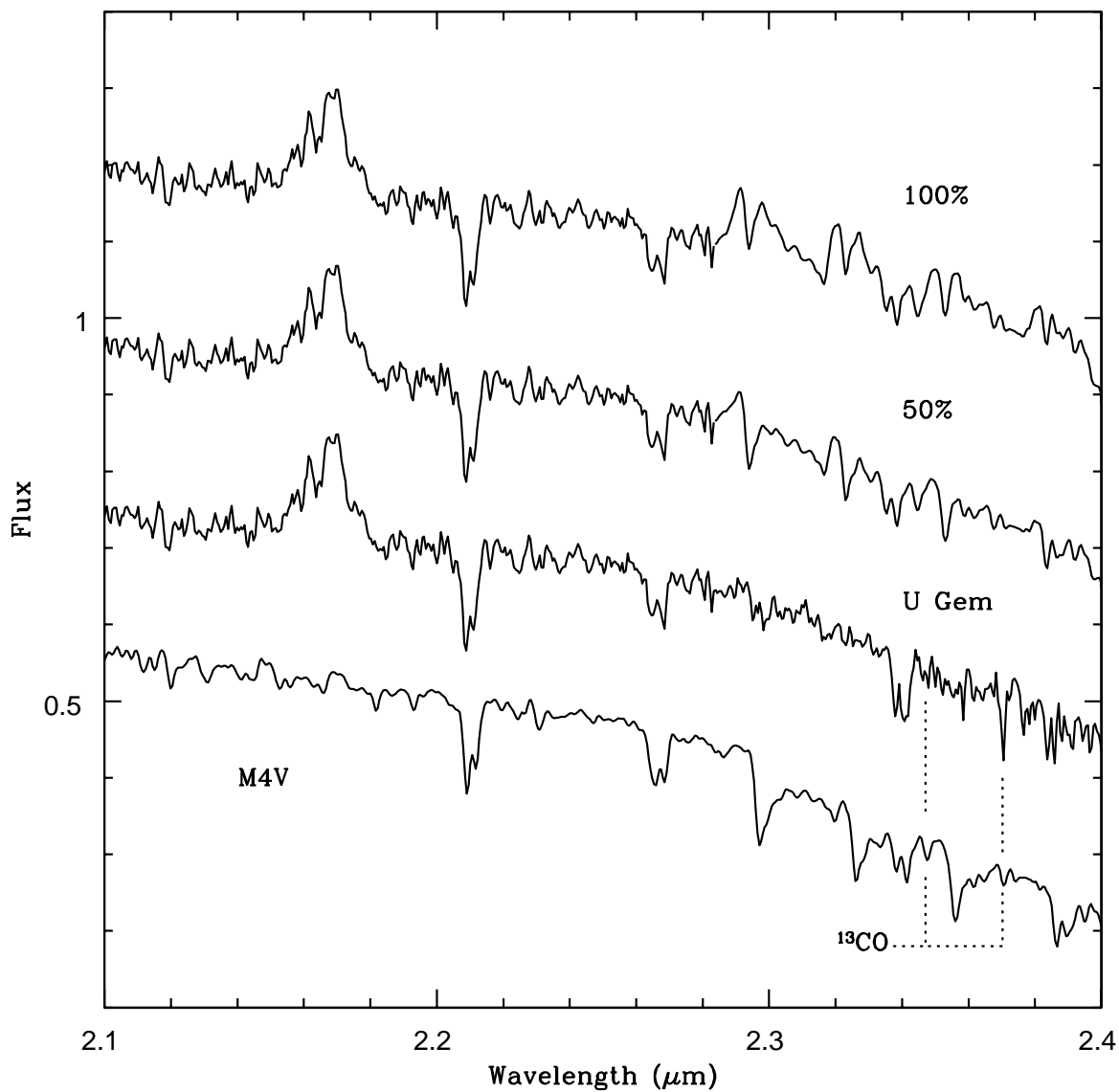}
\caption{ Model spectra used to assess CO emission contribution. Bottom spectrum is
a single star template M4V, the same spectral type as the secondary star in the dwarf nova U
Gem. The observed $K$-band spectrum of U Gem is shown as the second from the bottom, 
revealing its very weak CO absorption but apparently normal Na I and Ca I lines 
(see Harrison et al., 2005b). The top two spectra show our model M4V + CO emission spectrum 
combined with the observed U Gem spectrum. 
The model spectra show the choppy continuum and the trough created by the combination of 
the CO absorption from the secondary and the broader CO emission from the accretion disk. 
See text for details. The y-axis is normalized flux.
}
%\label{longlc}
\end{figure}

% ---------------------------------

\end{document}